\documentclass[%
 amsmath,amssymb,
 reprint,%
]{revtex4-1}

\usepackage{graphicx}
\usepackage{dcolumn}
\usepackage{bm}

\usepackage[utf8]{inputenc}
\usepackage[T1]{fontenc}
\usepackage{mathptmx}
\usepackage{etoolbox}

\makeatletter
\def\@email#1#2{%
 \endgroup
 \patchcmd{\titleblock@produce}
  {\frontmatter@RRAPformat}
  {\frontmatter@RRAPformat{\produce@RRAP{*#1\href{mailto:#2}{#2}}}\frontmatter@RRAPformat}
  {}{}
}%
\makeatother
\begin{document}

\title{High quality superconducting Nb co-planar resonators on sapphire substrate}
\author{S. Zhu$^{*}$}
 \email{szhu26@fnal.gov}
\author{F. Crisa}
\author{M. Bal}
\author{A. A. Murthy}
\author{J. Lee}
\author{Z. Sung}
\author{A. Lunin}
\author{D. Frolov}%
\author{R. Pilipenko}
\author{D. Bafia}
\author{A. Mitra}
\author{A. Romanenko}
\author{A. Grassellino}
\affiliation{Superconducting Quantum Materials and Systems Center (SQMS), \\Fermi National Accelerator Laboratory, \\Batavia, IL 60510, USA}



\begin{abstract}

We present measurements and simulations of superconducting Nb co-planar waveguide resonators on sapphire substrate down to millikelvin temperature range with different readout powers. In the high temperature regime, we demonstrate that the Nb film residual surface resistance is comparable to that observed in the ultra-high quality, bulk Nb 3D superconducting radio frequency cavities while the resonator quality is dominated by the BCS thermally excited quasiparticles. At low temperature both the resonator quality factor and frequency can be well explained using the two-level system models. Through the energy participation ratio simulations, we find that the two-level system loss tangent is $\sim 10^{-2}$, which agrees quite well with similar studies performed on the Nb 3D cavities (Refs.~\onlinecite{romanenko_understanding_2017, romanenko_three-dimensional_2020}).

\end{abstract}

\maketitle


It is well known that material losses play a key role in limiting the performance of superconducting circuits for the quantum computation ~\cite{oliver_materials_2013, wendin_quantum_2017, kjaergaard_superconducting_2020}. Among these losses, unpaired quasiparticles and two-level systems (TLSs) represent two major sources, and many efforts have been devoted to understand and minimize their contributions. Quasiparticles are typically generated in the superconducting film due to thermal excitation, cosmic radiation, or photon-phonon exchanging $\it{etc.}$ \cite{minev_planar_2013, wilen_correlated_2021, patel_phonon-mediated_2017}. They cause the energy dissipation, charge fluctuation, and correlated errors in the superconducting quantum devices \cite{martinis_saving_2021, christensen_anomalous_2019, wilen_correlated_2021, serniak_hot_2018}. TLSs represent a two-state defect common to the amorphous materials, which is seen as a major source of noise and decoherence in superconducting quantum devices. They parasitically couples to the applied field in the device and dissipate the energy into the environments \cite{muller_towards_2019, mcrae_materials_2020}. In 2D planar superconducting devices, the existence of amorphous oxides at the metal-air (MA), metal-substrate (MS), and substrate-air (SA) interfaces likely give rise to TLS \cite{gorgichuk_origin_2022}. Recent observations of hydride precipitates within niobium films may also be the host of TLSs \cite{murthy_tof-sims_2022}.


Due to numerous surfaces, interfaces, and defects present in the various material systems that comprise a superconductor, it can be difficult to directly pinpoint the impact a particular device region has on performance. To this end, superconducting resonators offer a simpler geometry that enables the systematic evaluation of various materials and interfaces. 
The development of high quality resonators also enables characterization of loss mechanisms on other materials in the quantum devices. For example, M. Checchin $\it{et.al.}$ used a 3D superconducting radio frequency (SRF) Nb cavity ($Q_i \sim 10^{11}$) to identify a new loss mechanism in the high-resistivity silicon substrates \cite{checchin_measurement_2022}. Therefore, accurate measurements on high quality superconducting resonators would allow for optimal selection of materials, device architectures, and circuit designs, in an effort to improve the coherence of a single quantum bit and enable the large scale of the circuit integration. Here we aim to identify the Nb film structures and its influence on the superconducting resonator quality.     

\begin{figure}
\includegraphics[scale=0.6]{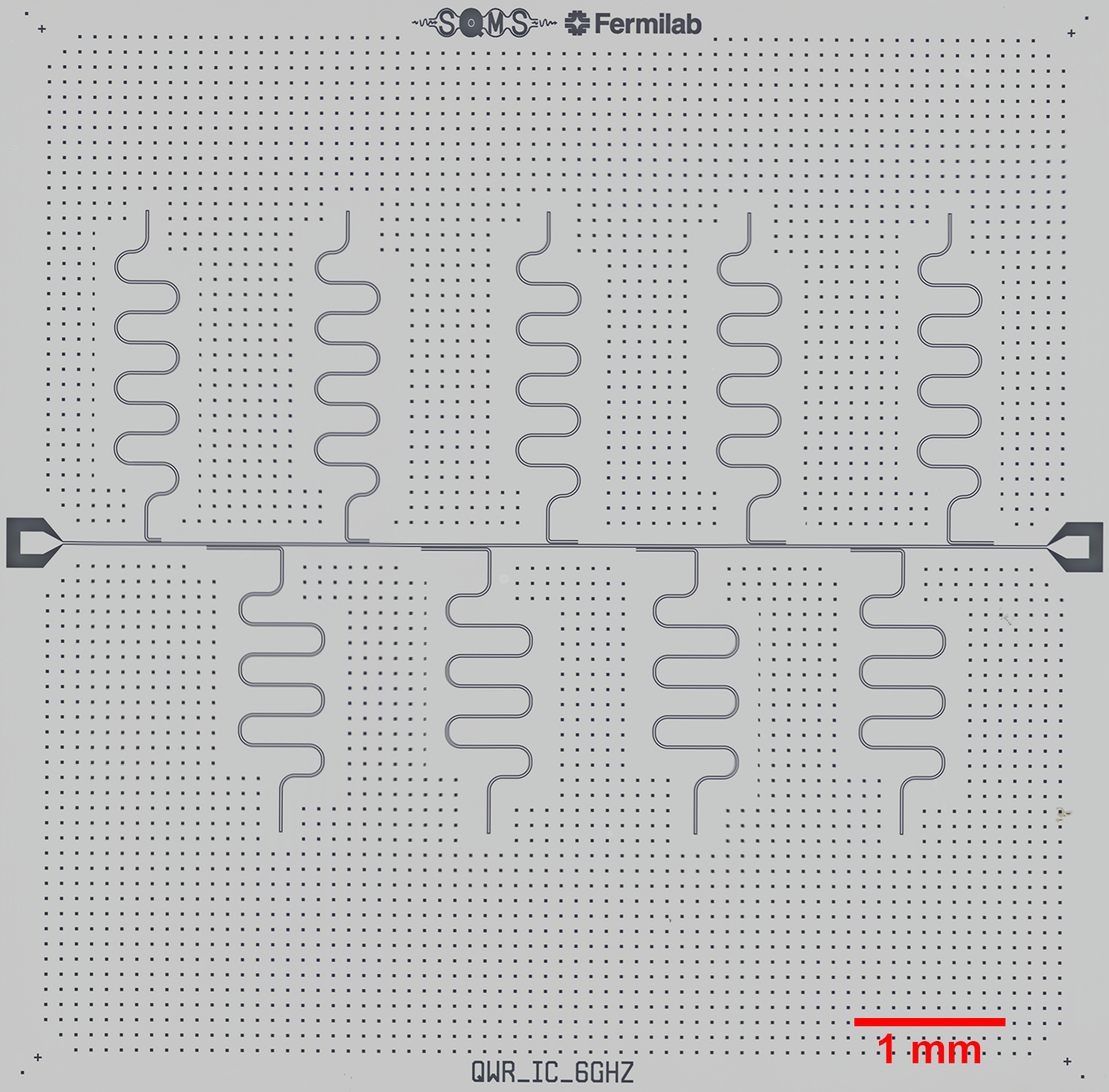}
\caption{\label{fig:epsart} Microscope of the device layout. Nine quarter-wave resonators inductively coupled (hanging on) a single $50 \Omega$ feedline.}
\end{figure}

In this work, 150 nm Nb film were sputtered on a c-axis sapphire substrate using DC magnetron sputtering technique. Prior to sputtering, the substrate surface was first solvent cleaned by applying ultrasonic power while immersing it sequentially in baths of heated NMP, IPA, Acetone and IPA. The solvent clean is followed by further cleaning  by submerging the substrate in a heated standard clean solution followed by DI water rinse and spin drying. The substrate is then immediately loaded into the  sputtering chamber withing, annealed at $200^\circ C$ for about 30 min and allowed to cool to room temperature under vacuum ( $\sim 4 \times 10^{-8}$ Torr) for 2 hours to remove the surface residuals and water vapors. The Nb film was patterned into co-planar waveguide (CPW) quarter wave ($\lambda/4$) resonators with a central trace width of $10\mu m$ and gap of $6\mu m$ by employing photolithography and fluorine based ICP reactive ion etching. On a $7.5\times7.5 mm^2$ chip, there are nine resonators inductively coupled to a $50\Omega$ feedline that the S21 transmission signals of all resonators can be measured using the vector network analyzer (see Fig. 1). The resonant frequencies are designed near 6GHz with a $\sim$ 50MHz step size and the coupling strength, $Q_{C}$, of the resonator to the feedline varies from 10k-100K by adjusting the coupler length. A separate 4-wire DC measurement on the patterned $20\mu m$ wide trace from the same wafer shows the Nb film superconducting transition temperature $T_C=9.3K$, from which the energy gap at zero temperature can be estimated to be $\Delta 0 = 1.76 k_B T_C = 1.41mV$. The room temperature sheet resistance is $R_{sq}=1.87\Omega /sq$, and combining with the measured $RRR=8.0$, the normal conductivity is given as $\sigma_n = 2.8 \times 10^7 S/m$. 


In order to assess the quality of the Nb film on the sapphire substrate, we carefully analyzed the film structure and chemical composition. As shown in Fig. 2a, we observe an abrupt interface between the Nb film and underlying sapphire substrate. There is minimal alloying present in this region, which differs from the alloyed regions previously reported for Nb films grown on silicon \cite{PhysRevMaterials.6.064402}. It is also apparent that the Nb film is textured and exhibits a single orientation from the electron backscatter diffraction orientation map (Fig. 2b). From the chemical analysis performed with time-of-flight secondary ion mass spectrometry (ToF-SIMS) (Fig. 2c),  we observe that impurities such as O, F, Cl, C, and H are all localized near the surface (0 nm). While F and Cl decay within the first few nanometers of the Nb film, O, C, and H persist throughout the entirety of the film and are likely present in the form of interstitial atoms, which is in agreement with previous reports. \cite{murthy_insights_2022, murthy_tof-sims_2022}.  

\begin{figure}
\includegraphics[scale=0.4]{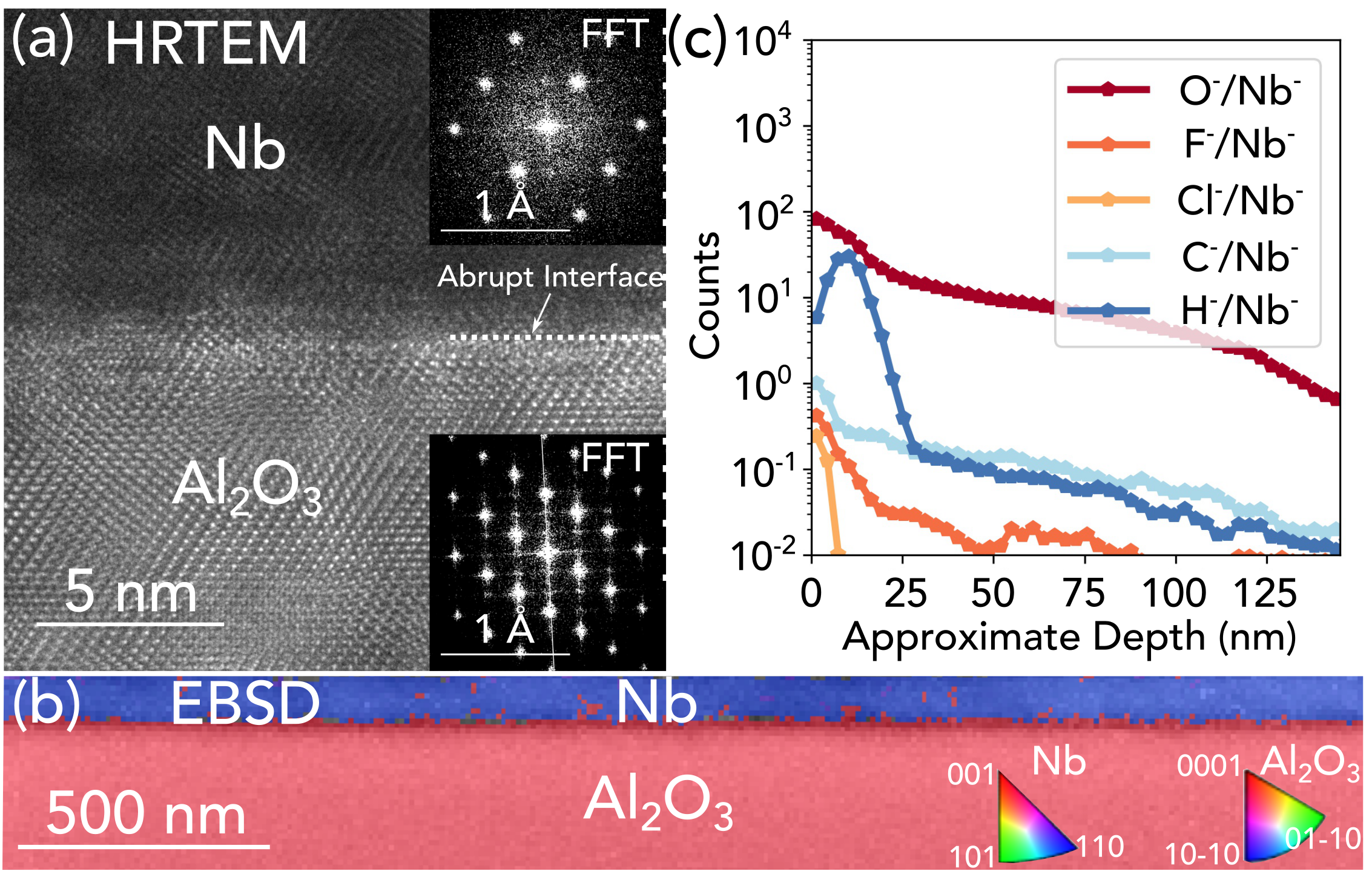}
\caption{\label{fig:epsart} Structural and chemical analysis of Nb film on Sapphire. (a) High resolution transmission electron microscopy (HRTEM) image of a cross-section taken of the interface between the Nb film and the underlying sapphire substrate. (b) Electron backscatter diffraction map of the same cross-section taken of the interface between the Nb film and the underlying substrate. (c) Time-of-Flight Secondary Ion Mass Spectrometry (ToF-SIMS) profiles taken from the Nb film provide information regarding the relative counts of various impurities as a function of depth.}
\end{figure}

The chip is packaged in a copper box coated with gold to improve the thermal contact with the mixing plate on the dilution refrigerator (DR). The intrinsic resonance of the box is simulated as 11GHz, which is specifically selected to eliminate interference with the device resonances. The feedline launchers are wire-bonded to the 50$\Omega$ printed circuit boards that are soldered to the non-magnetic SMA connectors. The whole enclosure is mounted inside a cryogenic $\mu$-metal can with IR filters to minimize losses due to the magnetic fluxes and stray IR radiations \cite{barends_minimizing_2011}. The wiring diagram is the same as described in Ref.~\onlinecite{mcrae_materials_2020}. A total 60dB attenuation ($\sim$70dB including the cable loss) at the input chain can effectively reduce the thermal photon radiation from the temperature instruments down below $10^{-3}$ at the mixing plate. A K$\&$L 0-12GHz low pass filter is used to prevent the high frequency noise from entering into the device. The microwave signal is amplified at 3K with a LNF high electron mobility transistor amplifier and a room temperature amplifier (both have gain $\sim 30dB$) before it is measured by a vector network analyzer. Two cascaded 4-8GHz cryogenic circulators are used to isolate the back actions from the amplifier to the resonators. 

Fig. 3(a) shows a typical measured S21 amplitude of one resonator on the chip. Using the method described by Mazin and Gao \cite{mazin_microwave_2021, gao_physics_nodate}, the total transmission $s_{21}$ through the measurement chain can be described by the equation
\begin{equation} 
s_{21}(f)=\alpha e^{-2\pi j f \tau}[1-\frac{Q_i/Q_c e^{j \phi_0}}{1+2jQ(\frac{f-f_r}{f_r})}],
\end{equation}
where complex constant $\alpha$ accounts for the gain/phase shift, $\tau$ the cable delay, and $\phi_0$ the circuit asymmetry. $Q_i$, $Q_c$, and $f_r$ are the resonator internal quality, coupling strength, and resonant frequency. $1/Q=1/Q_i+1/Q_c$ is the external quality. The resonator coupling strength presented in this paper is $Q_c = 5e4$. Using equation (1) we fit the transmission amplitude S21 as shown in Fig.3(a). We also plot the measurement (blue) and fits (red) in the complex I-Q plane (Fig. 3(b)) and evaluate the fitting confidence shown in Fig. 3(c) (here, residual is defined as the fitting data minus the measurement data). As we can see, the fitting residuals for both real and imaginary parts are less than $1 \%$, which is suggestive of a high confidence fitting.

\begin{figure}
\includegraphics[scale=0.4]{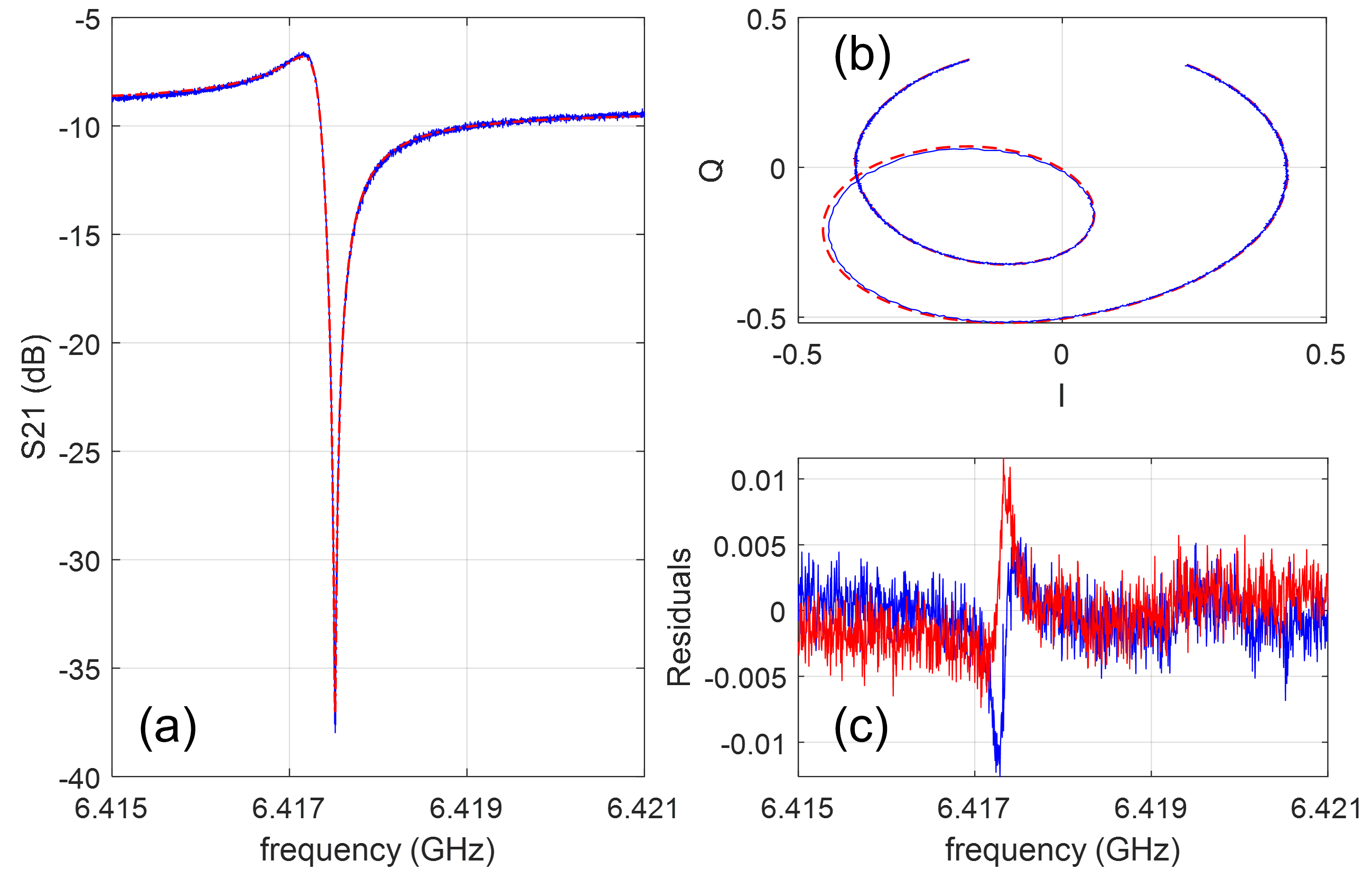}
\caption{\label{fig:epsart} Example of a fitting process to extract the resonator internal quality ($Q_i$), resonance ($f_r$), and coupling strength ($Q_c$). Blue lines are measured data and red lines the fitting curve in (a) and (b). Fig. c shows the residual of the real (blue) and imaginary (red) part between the fitting and experiment}
\end{figure}

We first measured the resonator $Q_i$ in the high temperature regime, where the quality of superconducting resonators is determined by the thermally excited quasiparticles that can be well described by Mattis-Barteen theory \cite{minev_planar_2013}. The complex conductivity $\sigma = \sigma_1-j\sigma_2$, which describes the response of both Cooper pairs and quasiparticles to a time varying electric field with $\hbar \omega < 2\Delta$, is given by:

\begin{subequations}
\label{eq:whole}
\begin{eqnarray}
\frac{\sigma_1}{\sigma_N}=&&\frac{1}{\hbar\omega}\{\int_{\Delta-\hbar \omega}^{-\Delta} g_1(1,2)tanh\frac{\hbar \omega + \epsilon}{2k_BT}d\epsilon +  \int_{\Delta}^{\infty} g_1(1,2)\nonumber\\
&&\times
(tanh\frac{\hbar \omega + \epsilon}{2k_BT}-tanh\frac{\epsilon}{2k_BT})d\epsilon\},
\end{eqnarray}

\begin{eqnarray}
\frac{\sigma_2}{\sigma_N} = &&\frac{1}{\hbar\omega}\{\int_{\Delta}^{\infty} [g_2(1,2)tanh\frac{\hbar \omega + \epsilon}{2k_BT}+g_2(2,1)tanh\frac{\epsilon}{2k_BT}]d\epsilon\nonumber\\ 
&&+
\int_{\Delta-\hbar \omega}^{\Delta} g_2(1,2)tanh\frac{\hbar \omega + \epsilon}{2k_BT}d\epsilon\}.    
\end{eqnarray}
\end{subequations}

Here $\sigma_N$ is the measured normal state conductivity. The coherence factors $g_{1,2}$ are given by
\begin{subequations}
\label{eq:whole}
\begin{equation}
g_1(1,2)=Re(N_1)Re(N_2)+Re(P_1)Re(P_2),
\end{equation}
\begin{equation}
g_2(1,2)=Im(N_1)Re(N_2)+Im(P_1)Re(P_2),
\end{equation}
\end{subequations}
where N and P are generalized densities of states of the superconductor and the arguments 1 and 2 represent the quasiparticle energy $\epsilon$ and $\epsilon + \hbar \omega$. Combining with the measured Nb film $T_C$ and energy gap $\Delta$, we can calculate the complex conductivity $\sigma$, and further get the surface impedance of the film $Z_s = R_s + jX_s = \sqrt{j\omega \mu_0 / \sigma}coth(d \sqrt{j\omega \mu_0 / \sigma})$, where $d=150nm$ is the measured film thickness.

The resonator loss is determined by the film surface resistance as $\delta = \frac{\alpha}{\omega \mu \lambda} R_s$ when it is dominated by the thermally excited quasiparticles, where $\alpha$ is defined as the ratio of the magnetic energy stored in the superconductor to the total magnetic energy, which is also known as the conductor participation ratio. It is equivalent to the geometry factor, $G$, commonly used in evaluating the SRF 3D Nb cavity for the accelerator application. $\lambda$ is the penetration depth and we take it as 80nm in this calculation. 

In Fig. 4, we plot the measured $Q_i$ (blue dot) as a function of the temperature, also the BCS fitting curve described above (green dash line). From the fitting we can extract $\alpha = 0.06$, a typical value for planar superconducting resonators ($\alpha \in [10^{-2}, 10^0]$) \cite{minev_planar_2013}. We also find that the BCS surface resistance at 1.5K is $R_s = 27n\Omega$. To compare the film quality with the SRF cavity that has the highest reported $Q \sim 10^{11}$ \cite{romanenko_three-dimensional_2020, romanenko_understanding_2017}, we convert the $R_s$ down to 1.3GHz (the BCS resistance scales with $\propto f^2$) and find $R_s=1.1n\Omega$, which is similar to that observed for the SRF cavity value ($R_s = 0.8n\Omega$). This relatively low surface resistance likely stems from the film being highly textured with a single orientation, as well as exhibiting minimal interfacial alloying and impurity content throughout the film (see Fig. 2).   

\begin{figure}
\includegraphics[scale=0.55]{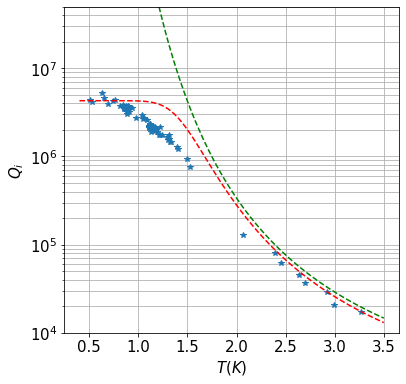}
\caption{\label{fig:epsart} Use BCS theory fitting (green dash line) to the measured $Q_i$ (blue star) to extract the conductor participation ratio $\alpha = 0.06$. Fitting with the total surface resistance $R_s = R_{res} + R_{BCS}$ to get the temperature independent residual resistance $R_{res} = 14 n\Omega$ (red dash line).}
\end{figure}

When the temperature is below 1.0K, the resonator quality reaches its upper limit and remains constant at $Q_i \sim 5 \times 10^6$. This can be explained by additional temperature-independent residual resistance, $R_{res}$, which arises from several sources such as random material defects or hydrides $\it{etc.}$. The total surface resistance is $R_s = R_{BCS} + R_{res}$. Using the same $\alpha$ value to fit the data (red dash line), $R_{res}$ is found to be $14n\Omega$, which,  after converting ($R_{res}=6.3n\Omega$ scales with $\sqrt{f}$), is larger than the SRF cavity value ($R_{res} \sim 2.2n\Omega$). A possible explanation of this larger residual resistance could be that the planar resonator has multiple interfaces (MA, SA, and MS), while in the SRF cavity there is only one residual layer at the cavity surface.

We also measured the resonator down to the DR base temperature ($T \sim 40mK$), where the TLS is believed to dominate the resonator loss. According to the model, the TLS induces both power- and temperature-dependent resonator loss, but only temperature-dependent resonance frequency shift \cite{gao_experimental_2008}. Therefore we can extract the TLS loss contribution from the resonator frequency measurement and avoid the power-dependent variation. Figure 5(a) shows the measured frequency shift $\Delta f/f_r = [f_{r}(T)-f_{r}(500mK)]/f_r$ as a function of temperature. Using the TLS model description,

\begin{equation}
    \frac{\Delta f}{f_r} = \frac{F \delta_{TLS}}{\pi}[Re(\Psi(\frac{1}{2} + \frac{\hbar \omega}{2 \pi j k_B T})) - log\frac{\hbar \omega}{2 \pi k_B T}],
\end{equation}


we can fit the measured data, as shown by the red dash line. In Eq. (4), $Re(\Psi)$ is the real part of the complex digamma function, and $\delta_{TLS}$ is the intrinsic TLS loss. The surface participation ratio, $F$, is defined as the ratio of total electrical energy stored in the TLS materials to the total electrical energy. Overall the data can be fitted quite well to the model; a minimum in $f_r$ around
$T = \frac{\hbar \omega}{2k_B} \sim 100mK $ 
meets the general TLS scenario \cite{muller_towards_2019}. The frequency shift is almost an order lower than those previously reported results \cite{gao_experimental_2008}. From fitting we can extract the production of $F\delta_{TLS} = 1.1 \times 10^{-6}$. Assuming the TLSs are hosted by the 5nm $Nb_2 O_5$ at the MA, SA, and MS interfaces with a dielectric constant $\epsilon = 33$, we can simulate the participation ratio $F=1 \times 10^{-4}$, therefore get $\delta_{TLS} \sim 10^{-2}$, which is close to the value from the 3D SRF cavity measurement \cite{romanenko_three-dimensional_2020, romanenko_understanding_2017}. In the simulation, we assume all the MA, SA, and MS interfaces are identical with same thickness and dielectric constant. However this may not be true in the actual geometry; more detailed analysis to clearly delineate the loss tangent of each layer is necessary \cite{calusine_analysis_2018}. 

Additionally we can extract the TLS loss  by fitting the resonator $Q_i$ as a function of the measurement power at fixed temperature: 
\begin{equation}
    \frac{1}{Q_i} = F \delta_{TLS} \frac{tanh(\hbar \omega / 2k_BT)}{\sqrt{1 + (\bar{n}/n_c)^{\beta}}}.
\end{equation}
Here $\bar{n}=P_{in}Q/\hbar \omega^2$ is the estimated average photon numbers injected into the resonator. Figure 5(b) shows the results of the measurement and fitting at $T=100mK$, from which we can extract the $F \delta_{TLS} \sim 2.8 \times 10^{-6}$, close to the fitting result from the frequency shift.

\begin{figure}
\includegraphics[scale=0.55]{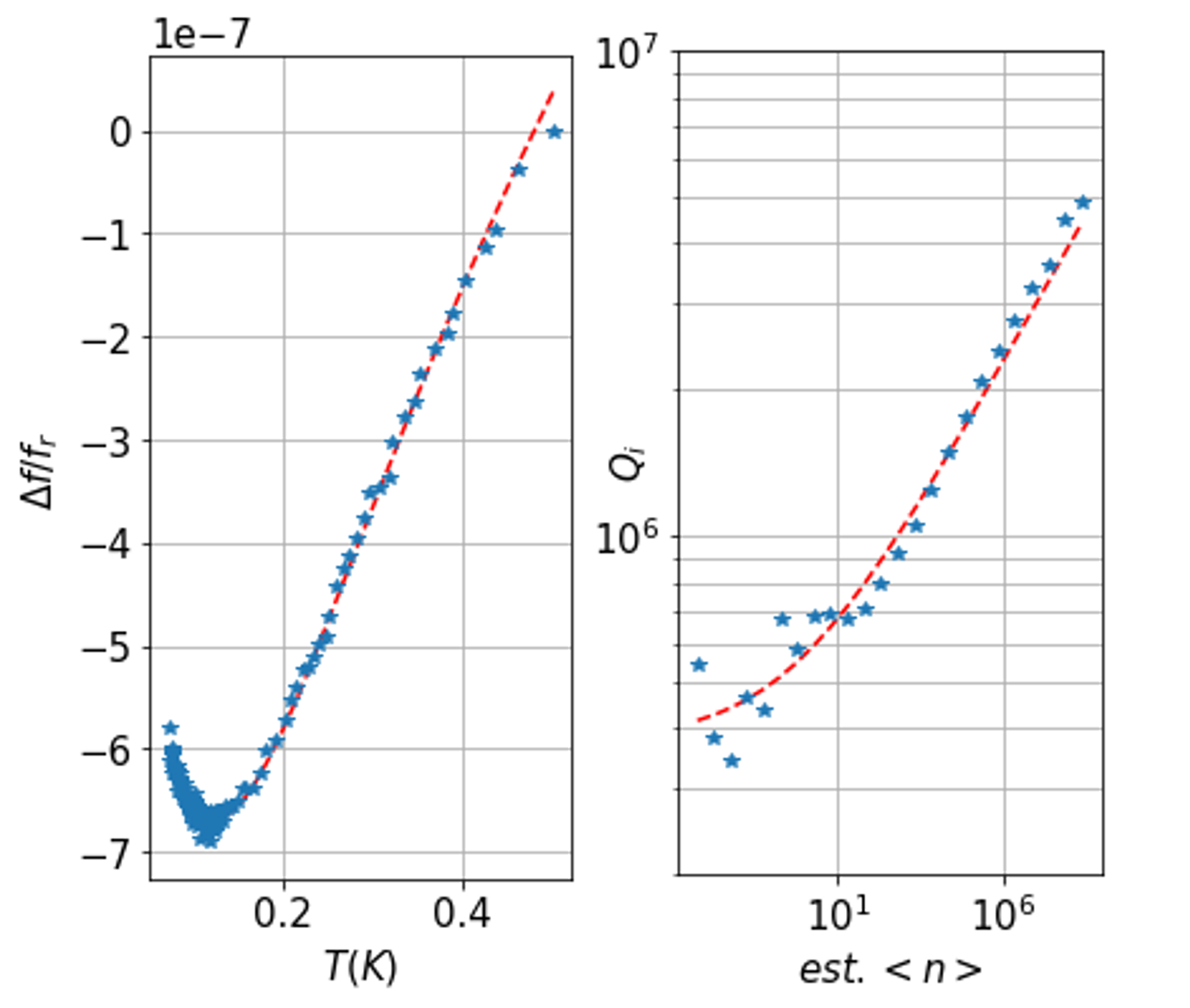}
\caption{\label{fig:epsart} Using the TLS model fitting to the resonator freqeuncy shift (left) and the resonator quality (right). Both fits give the similar TLS loss $F\delta_{TLS} \sim 10^{-6}$.}
\end{figure}

In conclusion, we carefully measured the Nb superconducting CPW resonator on the c-axis Sapphire substrate. We find that the BCS surface resistance of the Nb film is comparable to the Nb bulk used for the 3D SRF elliptical cavity. This is likely a result of the high structural and chemical quality indicated by various microscopy and mass spectrometry techniques. We also find that the residual resistance of the resonator is larger than what is typically observed in 3D Nb SRF cavities, which can be understood as an increase in the number of interface in the 2D planar resonator. At the low temperature regime, we find that the resonator is dominated by the TLSs that are possibly located in the native oxide $Nb_2O_5$ layers. By properly choosing the simulation parameters, we find the TLS loss to be $\sim 10^{-2}$, which is close to the reported value in the 3D cavity. Therefore we demonstrate that high quality Nb films can be prepared on the sapphire substrate.       

This material is based upon work supported by the U.S. Department of Energy, Office of Science, National Quantum Information Science Research Centers, Superconducting Quantum Materials and Systems Center (SQMS) under contract number DE-AC02-07CH11359. This work made use of the Pritzker Nanofabrication Facility part of the Pritzker School of Molecular Engineering at the University of Chicago, which receives support from Soft and Hybrid Nanotechnology Experimental (SHyNE) Resource (NSF ECCS-2025633), a node of the National Science Foundation’s National Nanotechnology Coordinated Infrastructure. This work made use of the EPIC facility of Northwestern University’s NU\textit{ANCE} Center, which has received support from the SHyNE Resource (NSF ECCS-2025633), the IIN, and Northwestern's MRSEC program (NSF DMR-1720139).

\nocite{*}
\bibliography{aipsamp}

\end{document}